\documentclass[]{agujournal2019}
\usepackage{url} %this package should fix any errors with URLs in refs.
\usepackage{amssymb}
\usepackage{pdfpages}
\usepackage{soul}

\draftfalse

\journalname{Geophysical Research Letters}

\begin{document}

\title{Phosphine in the Venusian Atmosphere: A Strict Upper Limit from SOFIA GREAT Observations}

%% ------------------------------------------------------------------------ %%
%
%  AUTHORS AND AFFILIATIONS
%
%% ------------------------------------------------------------------------ %%

% Authors are individuals who have significantly contributed to the
% research and preparation of the article. Group authors are allowed, if
% each author in the group is separately identified in an appendix.)

% List authors by first name or initial followed by last name and
% separated by commas. Use \affil{} to number affiliations, and
% \thanks{} for author notes.
% Additional author notes should be indicated with \thanks{} (for
% example, for current addresses).

% Example: \authors{A. B. Author\affil{1}\thanks{Current address, Antartica}, B. C. Author\affil{2,3}, and D. E.
% Author\affil{3,4}\thanks{Also funded by Monsanto.}}

\authors{M. A. Cordiner\affil{1,2}, G. L. Villanueva\affil{1}, H. Wiesemeyer\affil{3}, S. N. Milam\affil{1}, I. de Pater\affil{4}, A. Moullet\affil{5}, R. Aladro\affil{3}, C. A. Nixon\affil{1}, A. E. Thelen\affil{1}, S. B. Charnley\affil{1}, J. Stutzki\affil{3}, V. Kofman\affil{1}, S. Faggi\affil{1}, G. Liuzzi\affil{1,6}, R. Cosentino\affil{7}, B. A. McGuire\affil{8}}

%\authors{M. A. Cordiner\affil{1,2} et al.}

\affiliation{1}{Solar System Exploration Division, NASA Goddard Space Flight Center, 8800 Greenbelt Road, Greenbelt, MD 20771, USA.}
\affiliation{2}{Department of Physics, Catholic University of America, Washington, DC 20064, USA.}
\affiliation{3}{Max Planck Institute for Radio Astronomy, Auf dem H{\"u}gel 69, D-53121 Bonn, Germany.}
\affiliation{4}{Department of Astronomy, University of California, 501 Campbell Hall, Berkeley, CA 94720-3411, USA.}
\affiliation{5}{NASA Ames Research Center, Moffett Field, CA 94035, USA.}
\affiliation{6}{School of Engineering, Universit{\`a} degli Studi della Basilicata, 85100, Potenza, Italy}
\affiliation{7}{Space Telescope Science Institute, Baltimore, MD 21218, USA.}
\affiliation{8}{Department of Chemistry, Massachusetts Institute of Technology, Cambridge, MA 02139, USA.}

% \affiliation{3}{Third Affiliation}
% \affiliation{4}{Fourth Affiliation}

%% Corresponding Author:
% Corresponding author mailing address and e-mail address:

% (include name and email addresses of the corresponding author.  More
% than one corresponding author is allowed in this LaTeX file and for
% publication; but only one corresponding author is allowed in our
% editorial system.)

% Example: \correspondingauthor{First and Last Name}{email@address.edu}

\correspondingauthor{M. A. Cordiner}{martin.cordiner@nasa.gov}

%% Keypoints, final entry on title page.

%  List up to three key points (at least one is required)
%  Key Points summarize the main points and conclusions of the article
%  Each must be 140 characters or fewer with no special characters or punctuation and must be complete sentences

% Example:
% \begin{keypoints}
% \item	List up to three key points (at least one is required)
% \item	Key Points summarize the main points and conclusions of the article
% \item	Each must be 140 characters or fewer with no special characters or punctuation and must be complete sentences
% \end{keypoints}

\begin{keypoints}
\item A recent detection of phosphine (a possible biomarker) on Venus at millimeter wavelengths has been called into question by subsequent work.
\item We performed far-infrared spectroscopic observations of Venus using the GREAT instrument onboard the SOFIA aircraft.
\item Phosphine was not detected, and a strict upper limit on its atmospheric abundance of 0.8 ppb was derived between 75--110~km.
\end{keypoints}

%% ------------------------------------------------------------------------ %%
%
%  ABSTRACT and PLAIN LANGUAGE SUMMARY
%
% A good Abstract will begin with a short description of the problem
% being addressed, briefly describe the new data or analyses, then
% briefly states the main conclusion(s) and how they are supported and
% uncertainties.

% The Plain Language Summary should be written for a broad audience,
% including journalists and the science-interested public, that will not have 
% a background in your field.
%
% A Plain Language Summary is required in GRL, JGR: Planets, JGR: Biogeosciences,
% JGR: Oceans, G-Cubed, Reviews of Geophysics, and JAMES.
% see http://sharingscience.agu.org/creating-plain-language-summary/)
%
%% ------------------------------------------------------------------------ %%

%% \begin{abstract} starts the second page

\begin{abstract}
The presence of phosphine (PH$_3$) in the atmosphere of Venus was reported by \citeA{gre21a}, based
on observations of the $J=$1--0 transition at 267~GHz using ground-based, millimeter-wave spectroscopy. This unexpected discovery presents a challenge for our understanding of Venus's atmosphere, and has led to a reappraisal of the possible sources and sinks of atmospheric phosphorous-bearing gases. Here we present results from a search for PH$_3$ on Venus using the GREAT instrument aboard the SOFIA aircraft, over three flights conducted in November 2021. Multiple PH$_3$ transitions were targeted at frequencies centered on 533~GHz and 1067~GHz, but no evidence for atmospheric PH$_3$ was detected. Through radiative transfer modeling, we derived a disk-averaged upper limit on the PH$_3$ abundance of 0.8 ppb in the altitude range 75--110~km, which is more stringent than previous ground-based studies.
\end{abstract}

%\section*{Plain Language Summary}

%We performed observations using a unique, flying telescope --- the Stratospheric Observatory for Infrared Astronomy (SOFIA) --- to search for a gas called phosphine in the atmosphere of Venus, which has been suggested to be an indicator for life. Our observations showed no evidence of phosphine, in contrast to the original claim, {and we provide a stringent upper limit on its abundance. Two} other recent studies using different telescopes also found no evidence for phosphine on Venus, in agreement with our present result. 

\section{Introduction}
\begin{figure}
\centering
\includegraphics[width=0.55\textwidth]{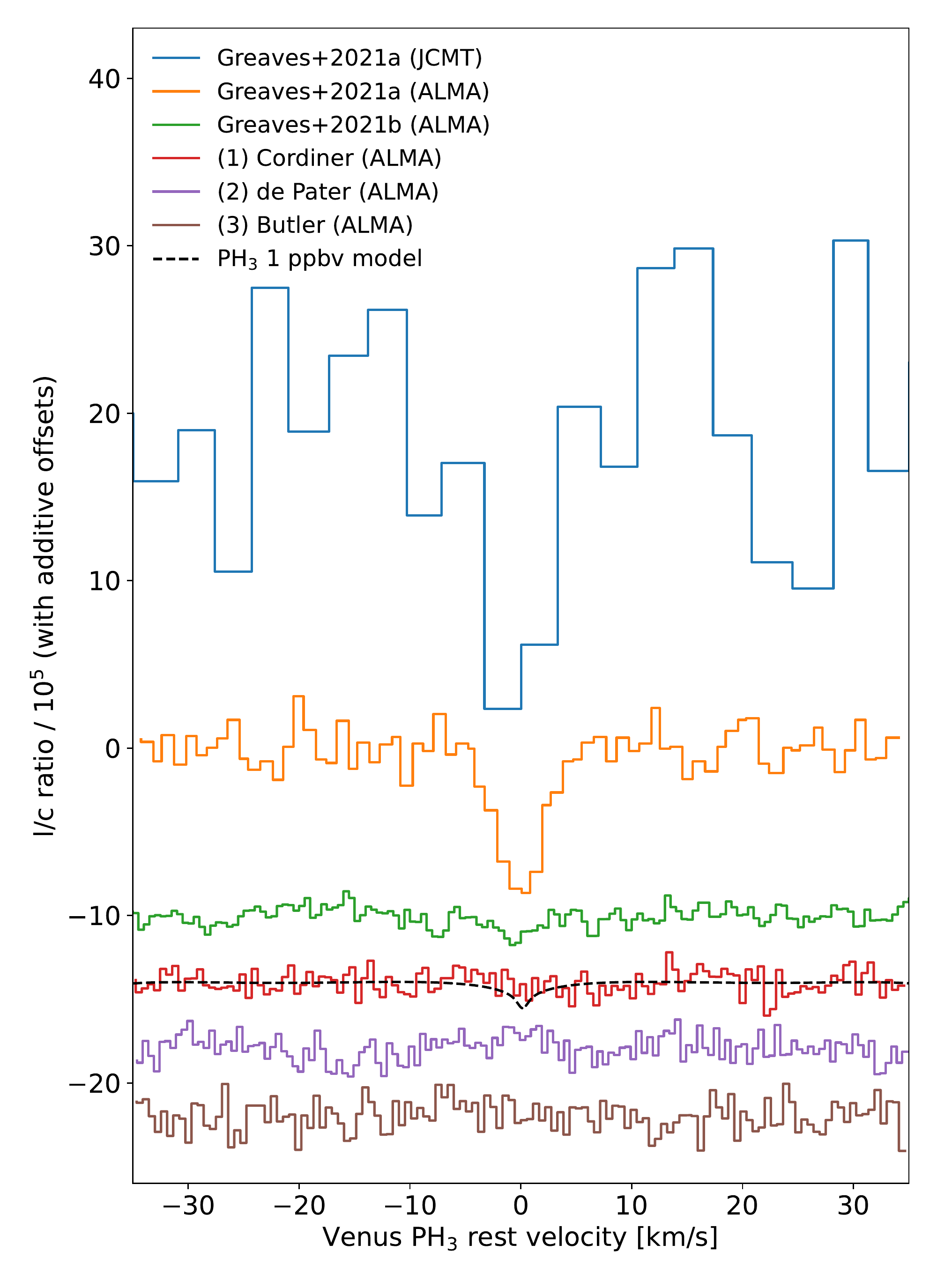}
\caption{JCMT and ALMA spectra from the original Venus study of \citeA{gre21a} (top two traces), showing the spectral region in the vicinity of the PH$_3$ $J=1$--0 line at 266.945~GHz. The lower four traces show disk-averaged ALMA spectra derived from the same data using different calibration and reduction methodologies by different authors, as explained by \citeA{vil21} and \citeA{gre21b}. For reference, the black dashed trace shows the 1~ppb PH$_3$ spectral model (upper limit) from \citeA{vil21}.  {The Greaves et al. (2021a) and (2021b) ALMA spectra differ considerably, due to improper calibration of the 2021a data.}\label{fig:intro}}
\end{figure}

\citeA{gre21a} presented the first evidence for phosphine (PH$_3$) gas on Venus, based on a tentative detection of the $J=1$--0 line at 267~GHz using the James Clerk Maxwell Telescope (JCMT), in addition to a seemingly more convincing detection of the same line using the Atacama Large Millimeter/submillimeter Array (ALMA), at a formally-derived statistical significance of $15\sigma$ (Figure \ref{fig:intro}). {A disk-averaged PH$_3$ abundance of 20 ppb was obtained by \citeA{gre21a}, but following recalibration of the ALMA data, a smaller value in the range $\sim1$--7~ppb was derived \cite{gre21b}}. The claimed detection of PH$_3$ was surprising considering the difficulty of producing this gas from abiotic mechanisms such as atmospheric photochemistry, geochemistry or meteorological processes \cite{gre21a,bai20}, which led to speculation regarding a possible biological origin. The potential utility of PH$_3$ as a biomarker (indicator for the presence of life) in planetary and exoplanetary atmospheres is presently under investigation \cite{sou20,wun21}. However, theorizing and speculation regarding the origin of Venusian PH$_3$ at parts-per-billion levels remains premature, considering several problems have been identified with the \citeA{gre21a} data calibration and analysis procedures by multiple, independent studies \cite{sne20,vil21,tho21,lin21,aki21}. In contrast to \citeA{gre21a}, these subsequent investigations found no significant evidence for PH$_3$ absorption in the same JCMT and ALMA data. The tentative JCMT detection can be explained as a statistical outlier, amplified/introduced by repeated subtraction of fitted, periodic baseline ripples, whereas the significance of the initial ALMA detection is reduced to (at or near) zero when conventional bandpass calibration procedures are applied. It was further shown by \citeA{vil21} that the presence (or not) of a putative absorption feature at the PH$_3$ $J=1$--0 frequency is dependent on the order of polynomial used in the baseline subtraction, making it difficult to obtain a reliable PH$_3$ detection from those data.

The JCMT and ALMA (disk-averaged) spectra from \citeA{gre21a} and \citeA{gre21b} are compared with three different, standard, reduction methods for the same ALMA data in Figure \ref{fig:intro} (see \citeA{vil21} for further details). The Venus atmospheric model for a PH$_3$ abundance of 1~ppb (the upper limit derived by \citeA{vil21}) is also overlaid on top of the red curve.

{Follow-up} attempts to detect PH$_3$ on Venus using infrared spectroscopy were performed using {archival data from} the NASA Infrared Telescope Facility (IRTF) on Hawaii \cite{enc20}, and from the Venus Express orbiter \cite{tro21}. These studies were sensitive to PH$_3$ in the altitude range $\sim60$--95~km, but no phosphine was detected. On the other hand, a reanalysis of Pioneer Venus mass spectrometry data by \citeA{mog21} revealed peaks at masses (per unit charge) of 33.997 amu/q and 32.989 amu/q, consistent with the presence of PH$_3^+$ and PH$_2^+$, respectively, at an altitude of 51.3~km. However, the close overlap of these mass peaks with H$_2$S$^+$ (33.988 amu) and HS$^+$ (32.980) introduces ambiguity into these molecular detections. Furthermore, the detection of (ionized) atomic P$^+$ by \citeA{mog21} was of marginal significance, considering the error bars. Meanwhile, further analysis by \citeA{gre22} of the original ALMA and JCMT data continues to support the case for PH$_3$ being present in the atmosphere of Venus. 

To help resolve the debate concerning the possible presence of phosphine on Venus, we performed spectroscopic observations using the Stratospheric Observatory for Infrared Astronomy (SOFIA) targeting multiple rotational transitions of PH$_3$ at high sensitivity and high spectral resolution. Due to its operating altitude of around 13~km, SOFIA has unique access to high-frequency submillimeter and infrared emission, which is not accessible using ground-based astronomy due to the opacity of Earth's atmosphere at these wavelengths. Our aim was to detect a set of the stronger, higher-energy rotational transitions of collisionally-excited PH$_3$, to augment prior observations and provide a robust measure of the PH$_3$ abundance in Venus's atmosphere. By virtue of the extremely high spectral resolution of the SOFIA heterodyne instrument (up to $R=4\times10^6$ at 1067~GHz), our observations provide altitudinal sensitivity for phosphine in the range 75--110~km, extending to higher altitudes than \citeA{enc20} and \citeA{tro21}, with global coverage of Venus. Our new study therefore provides spatial coverage that is highly complementary to the PH$_3$ upper limits obtained previously at shorter infrared wavelengths.

\section{Materials and Methods}

\subsection{Observations}
\label{sec:obs}

Observations were performed using the German REceiver for Astronomy at Terahertz Frequencies (GREAT), on the SOFIA telescope, over three flights conducted on UT 2021-11-10, 12 and 13 as part of Director's discretionary time program \#75\_0059. During this time, Venus was at a geocentric distance of 0.58~au, with an angular diameter of $28''$ and a solar illumination phase of $98^{\circ}$--$100^{\circ}$. The total integration time on Venus was 36 minutes.

GREAT is a modular-design heterodyne spectrometer operating at submillimeter and far-infrared wavelengths. For this experiment, the ``4GREAT'' instrument setup was used \cite{dur21}, consisting of four single-pixel mixers, out of which the first two units --- 4G1 and 4G2 (carrying superconductor-insulator-superconductor mixers {and operating at frequencies around 0.5 and 1.0 THz, respectively}) --- were used to target phosphine.

Typical single-sideband receiver temperatures for 4G1 and 4G2 were 230 K and $\sim1000$ K, respectively. The zenith precipitable water vapor column was close to 0.02~mm for all three flights, and the corresponding zenith opacities were $\sim0.12$ in both bands. The mean aircraft altitude was 12.2~km. Total system temperatures were 350--370 K for 4G1 and 1100--1570 K for 4G2, increasing during each flight due to the decreasing elevation angle of Venus (typically from $21^{\circ}$ to $16^{\circ}$).

The total power signal due to Earth's atmosphere was removed from the spectra by fast-chopping (at 0.6 Hz), alternating between two opposing reference positions $3'$ away from the science target. Telescope guiding was achieved by targeting Venus's opto-center, which resulted in an $8''$ offset in the science pointing position away from the geometric center of the planet's disk. The raw spectroscopic data were calibrated to Rayleigh-Jeans equivalent brightness temperatures using calibration loads at ambient and cold temperatures ($\approx300$~K and 68 K, respectively).

The extended fast Fourier transform spectrometers \cite{kle12} cover an intermediate-frequency (IF) base-band ranging from 4 to 8 GHz, utilizing 16,384 frequency channels, resulting in a generic spectral resolution of 244 kHz. The 4G1 and 4G2 local oscillator frequencies were configured to simultaneously observe the PH$_3$ $J=2-1$ doublet at 534 GHz (in the 4G1 lower sideband; LSB) and the PH$_3$ $J=4-3$ quadruplet at 1067~GHz (in the 4G2 upper sideband; USB). The SOFIA beam FWHM was $54''$ for the 4G1 channel and $25''$ for 4G2.  The spectral axis was Doppler-corrected to the Venus rest frame, and later rebinned by a factor of four (to a spectral resolution of 976~kHz) for subsequent data analysis. Individual line frequencies and basic spectroscopic parameters are given in Table S1, including the Einstein $A$ coefficient, the upper-state statistical weight ($g_u$), and the upper-state energy ($E_u$) of each transition.

\subsection{Telluric Modeling and Continuum Fringe Removal}
\label{sec:defringing}

\begin{figure}
\centering
\includegraphics[width=\columnwidth]{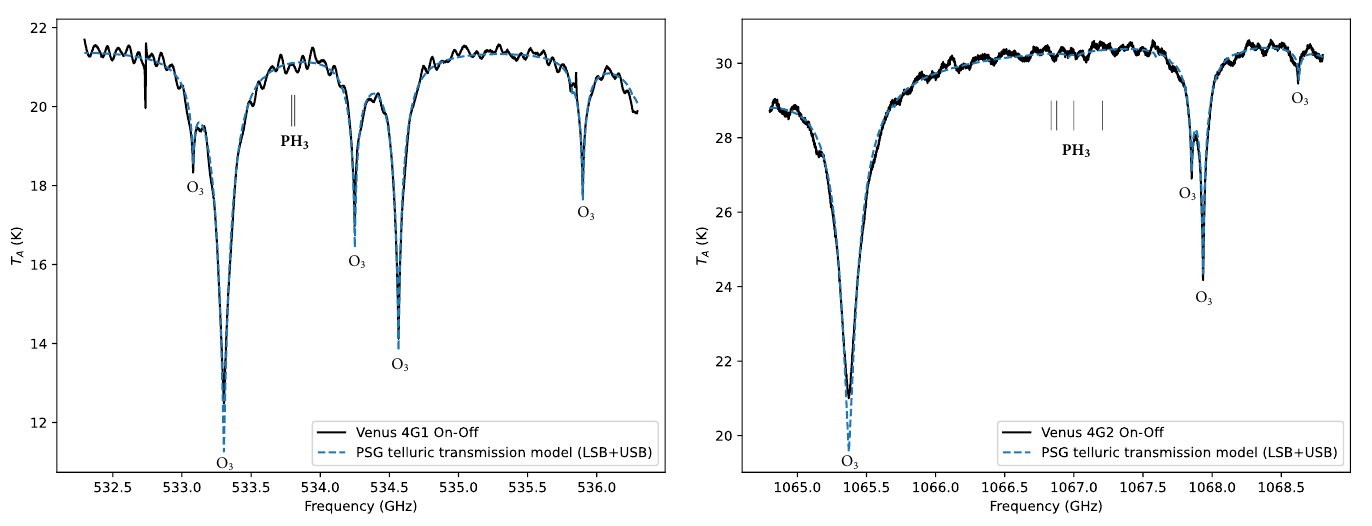} 
\caption{SOFIA Venus spectra from the 4G1 (left) and 4G2 (right) channels of the GREAT receiver. PSG terrestrial transmission models are overlaid, highlighting the presence of strong telluric O$_3$ absorptions. The spectra contain overlapping contributions from the lower and upper receiver sidebands. Vertical bars indicate PH$_3$ line frequencies. \label{fig:wideSpectra}}
\end{figure}

The observed Venus spectra (Figure \ref{fig:wideSpectra}), are composed of a linear combination of contributions from the upper and lower receiver sidebands, in a ratio of approximately $1:1$. The resulting spectra are dominated by the Venus thermal continuum, which is overlaid by several deep absorption lines due to terrestrial (telluric) ozone (O$_3$). A quasi-periodic fringe pattern is also evident in the continuum of both the 4G1 and 4G2 bands, with an amplitude of about 0.7\% of the continuum level, and an appearance characteristic of multiple overlapping sine waves of different frequencies. These fringes arise primarily due to internal reflections between the hot and cold calibration loads and the SOFIA subreflector, further modulated by the interfering contributions from the two sidebands (LSB + USB) in each receiver channel, and are therefore nontrivial to characterize {(see Supporting Information text, Section S1)}. 

Fourier analysis is a valid approach for identifying and removing such periodic spectral components, but for the present study, we adopt an iterative Lomb-Scargle `periodogram' method \cite{sca82}. This involves cross-correlating the observed spectra with a set of sine waves spanning a broad, quasi-continuous range of angular frequencies ($\omega=2\pi/T_{\nu}$, where $T_{\nu}$ is the wave period in frequency space). The peak cross-correlation amplitude is then plotted as a function of $\omega$ to produce the periodogram. Specific $\omega$ values were identified from the global maxima of the periodogram, and their corresponding sine waves were fitted one-by-one to the observed spectrum, optimizing the wave phases and amplitudes using least-squares minimization. The individual waves were subtracted sequentially from the observed spectra, and the periodogram was regenerated after each iteration. Our iterative fringe removal procedure was terminated when no further clear periodic signals could be identified in the data, which occurred after seven iterations for both the 4G1 and 4G2 channels (see Supporting Information Figures S1 and S2).

The Lomb-Scargle periodigram method provides a more nuanced approach than Fourier analysis, since it identifies and removes a single, fully characterized sine-wave component at each step, whereas Fourier analysis tends to be more destructive, removing a broader range of harmonics, with poorly-characterized phases and amplitudes. The Lomb-Scargle method also works with masked data, which helps prevent accidental removal of real spectral features. Furthermore, an identical set of wave components can be removed from the observations and from the spectral models for Venus, which allows investigation of the degree to which various (real) atmospheric signals may be corrupted as a result of continuum fringe removal.

For optimal removal of the instrumental fringes using this method, it is necessary to work with a baseline-subtracted spectrum, with the continuum and telluric lines removed. The fitted (continuum + telluric) model is later added back to the de-fringed spectra to facilitate subsequent analysis.  A spectral model for Venus (as viewed through Earth's atmosphere), was therefore constructed using the Planetary Spectrum Generator (PSG) \cite{vil18}, following the same methodology as \citeA{vil21}, {using gas abundance data from \citeA{ehr12} and the VIRA-45 temperature profile of \citeA{zas06}}. Our model geometry specifically accounts for the size and offset of the SOFIA beam with respect to Venus {(see Section \ref{sec:obs})}. The resulting model spectra for the upper and lower sidebands of the 4G1 and 4G2 channels were multiplied by corresponding PSG models for the terrestrial transmittance in the direction of Venus at the time of our SOFIA observations, taking into account the aircraft's mean latitude, longitude, altitude, and velocity along the line of sight. For each 4GREAT channel, the two model spectra for each sideband were overlaid and averaged together. A uniform scaling factor of 0.72 was applied to the terrestrial O$_3$ vertical abundance profile to obtain the best fit to the observed spectra, shown with dashed blue curves in Figure \ref{fig:wideSpectra}. A linear (multiplicative) gain slope was also included in the spectral model fits, to account for the varying 4GREAT amplitude response as a function of frequency. 

During fitting of the continuum, as well as for subsequent removal of the instrumental fringes, the spectral regions within $\pm5$~km\,s$^{-1}$ of the PH$_3$ lines were masked, therefore avoiding biasing the fit as a result of the presence (or absence) of PH$_3$, which was not included in the Venus atmospheric model at this stage. Due to the severity of telluric O$_3$ absorption in our spectra, and the difficulty in obtaining a perfect fit to the cores of the deepest telluric lines, we chose to perform all subsequent analysis (including fringe removal) on a restricted frequency range of $\pm100$~km\,s$^{-1}$ either side of the PH$_3$ lines (533.4--534.1~GHz for 4G1 and 1066.3--1067.7~GHz for 4G2), where the continuum fit was sufficiently good. 

\section{Results}

\begin{figure}
\centering
\includegraphics[width=\columnwidth]{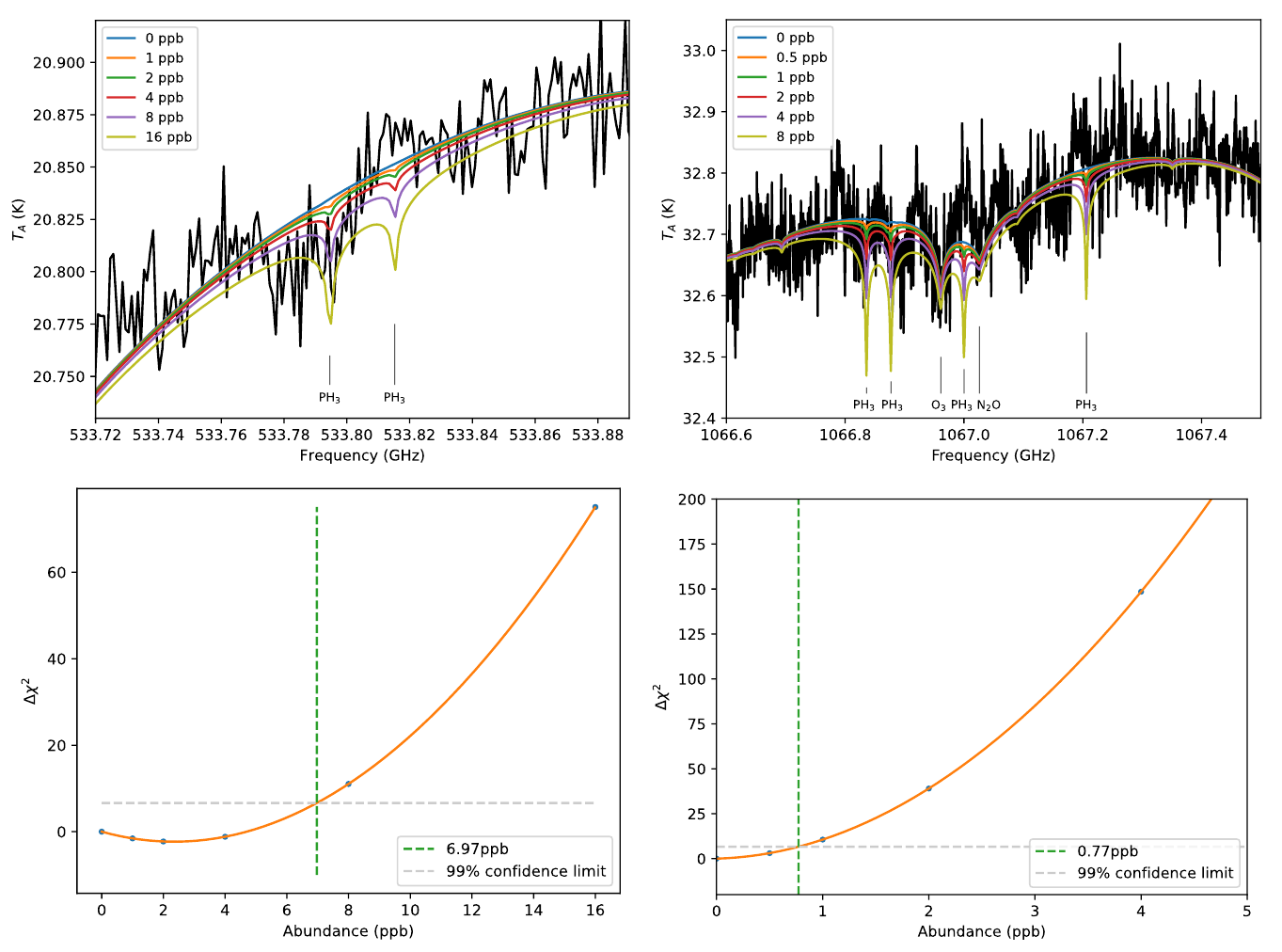} 
\caption{SOFIA GREAT Venus spectra in the 4G1 (top left) and 4G2 (top right) channels (shown in black). A family of PSG models is overlaid for each channel (coloured curves), spanning a range of different PH$_3$ abundances as identified in the legend. Major spectral features are labeled. Lower panels show the corresponding $\Delta\chi^2$ curve for each channel; the abundances corresponding to the 99\% confidence limit (\emph{i.e.} $\Delta\chi^2=6.63$) are highlighted with green dashed lines. \label{fig:results}}
\end{figure}

Observed, de-fringed SOFIA spectra in the vicinity of the targeted PH$_3$ lines are shown in Figure \ref{fig:results}. The RMS noise is 24 mK for the 4G1 spectra and 59 mK for 4G2, which is reasonably close to the theoretical values of 14~mK and 50~mK, respectively. The locations of the PH$_3$ lines are labeled, and for the 4G2 channel, broad contaminating lines due to telluric O$_3$ and NO$_2$ are also identified. These contaminants originate at rest frequencies of 1078.686~GHz and 1078.621~GHz in the upper 4G2 sideband, and are present in both the observed and modeled spectra. Several other, weaker lines due to terrestrial O$_3$ contamination are visible in the 4G2 model spectra, but these have not been labeled. No clearly-identifiable PH$_3$ lines can be seen in either the 4G1 or the 4G2 frequency ranges.

We performed a statistical analysis to quantify the amount of PH$_3$ that could be present in Venus's atmosphere, given the noise level of our data. A set of spectral models was generated as in Section \ref{sec:defringing}, consisting of Venus + telluric spectra for the LSB and USB in each 4GREAT channel, including (for Venus) CO$_2$, N$_2$, CO, O$_2$, SO$_2$, H$_2$O, O$_3$, He and PH$_3$ gases. Starting with a PH$_3$ abundance of zero, and with the $\pm5$~km\,s$^{-1}$ regions surrounding the PH$_3$ lines masked, the model Venus continuum was scaled to obtain the best fit to the observations. Using a constant PH$_3$ abundance with altitude, the amount of PH$_3$ in the model was steadily increased, and the difference between the model and observations was monitored using the $\chi^2$ statistic (\emph{e.g.} \citeA{nix10}). The change in $\chi^2$ as a function of the PH$_3$ abundance $a$ {($\Delta{\chi^2}(a)=\chi^2(a)-\chi^2_0$, where $\chi^2_0$ is the chi-square value at $a=0$)} is plotted for 4G1 and 4G2 in the lower panels of Figure \ref{fig:results}.

For 4G1, a minimum in the chi-square curve occurs at a PH$_3$ abundance of $a=2.3$~ppb, for which $\Delta\chi^2$ is $-2.3$. This corresponds to only a $1.5\sigma$ confidence interval, and is therefore not a significant detection. The 99\% statistical confidence limit is reached for $a=6.97$~ppb, which we consider to be a robust upper limit on the PH$_3$ abundance for this (4G1) channel. The PH$_3$ lines in the 4G2 channel provide a more sensitive upper limit than 4G1 due to their larger Einstein $A$ values and statistical weights (Table S1), in addition to higher energy level populations in the atmospheric regions we are probing. The $\Delta\chi^2$ curve for 4G2 is at a minimum for $a=0$ and rises to reach the 99\% confidence threshold at $a=0.77$~ppb. This represents a stringent upper limit on the average PH$_3$ abundance within the 4G2 beam.

The 4G1 PH$_3$ upper limit represents a robust disk-averaged measurement due to the relatively large beam size ($54''$) compared with Venus's $28''$ disk. On the other hand, the smaller ($25''$) 4G2 beam --- offset $8''$ from Venus's disk-center --- integrates only 56\% of the total planetary disk emission (assuming spatially-uniform emission), and primarily samples dayside (afternoon) equatorial latitudes. Adopting the low-latitude ($\phi<35^{\circ}$) VIRA temperature profile for the Venusian afternoon (solar time 13:30--18:00) from \citeA{zas06}, results in a 4\% increase in the modeled beam-averaged PH$_3$ ($J=4-3$) line depths relative to the continuum. Consequently, the PH$_3$ upper limit from 4G2 may be further reduced to $a=0.74$~ppb. In reality, due to the Gaussian beam shape (with its broad wings extending over the entire planet), our observations are sensitive to a somewhat broader range of latitudes and solar times, so the true upper limit should be in the range $a=0.74$--0.77~ppb. An additional error margin of a few per-cent should also be included, considering uncertainties in the adopted VIRA temperature profiles. A more accurate upper limit would therefore be obtained using actual temperature measurements for the global Venusian atmosphere, contemporaneous with our observations, but in the absence of such data, we conservatively round the PH$_3$ abundance upper limit up to 0.8~ppb.

\section{Discussion}

Our upper limit represents confirmation of the $\sim1$~ppb upper limit derived by \citeA{vil21} using the \citeA{gre21a} ALMA data (based on a spatially-filtered average of the Venusian disk), but is at odds with the 20~ppb derived by \citeA{gre21a} using JCMT spectra. Although we cannot rule out temporal variability of the putative PH$_3$ signal, our result is consistent with the findings of \citeA{tho21}, who determined that the PH$_3$ absorption identified by \citeA{gre21a} in their JCMT spectra may have been an artifact resulting from high-order polynomial baseline correction.

Our SOFIA data also suffer from significant baseline ripples (fringes). It is therefore important to evaluate the degree to which our results may have been affected by our fringe removal strategy. Masking the PH$_3$ lines of interest during fringe removal was a crucial initial step to help avoid accidental fitting and removal of any real PH$_3$ signal. The masked regions ($\pm5$~km\,s$^{-1}$; 18~MHz either side of each 4G2 PH$_3$ line) were selected to mark the approximate boundary for where the pressure-broadened model PH$_3$ lines become difficult to distinguish from the continuum. Care was also taken to ensure these regions were narrow enough to leave enough continuum visible for reliable periodogram generation (see Figures S1 and S2).   

Figure S3 shows a demonstration that our defringing method has no significant impact on the strengths of the narrow parts of the PH$_3$ lines (within the $\pm5$~km\,s$^{-1}$ masked regions). These figures were generated by fitting and removing from the models in sequence, the same set of periodic waves as identified for the observed spectra (Section \ref{sec:defringing}), by varying their amplitudes but keeping their phases and $\omega$ values fixed. At each step, the best-fitting sine wave was subtracted until all seven frequency components were removed. Some minor distortion of the broad line wings and continuum is evident, {but this is well below the noise level (as shown in the lower panels of Figure S3)}. Consequently, our abundance analysis is expected to be reliable for absorption originating from the narrow PH$_3$ line core regions.

\begin{figure}
\centering
\includegraphics[height=2.5in]{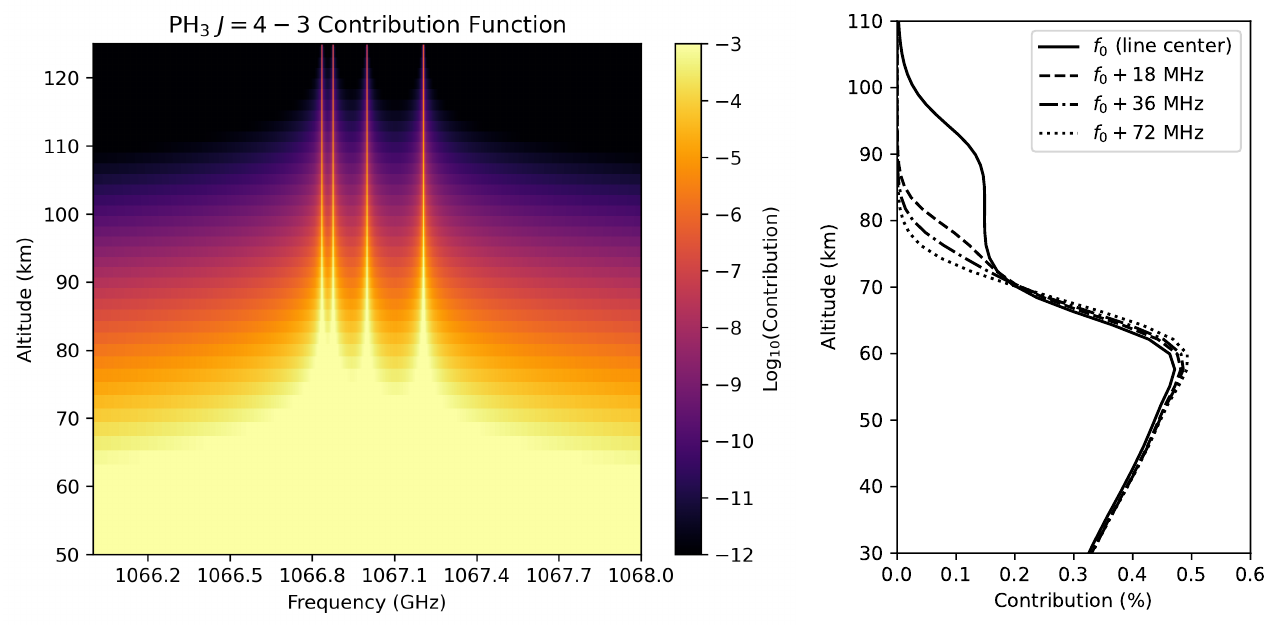} 
\caption{Left panel: contribution function from our Venus radiative transfer model, showing the fraction of the total spectral opacity as a function of frequency and altitude for the 4G2 band. The four prominent peaks are due to the PH$_3$ $J_K=4_K-3_K$ lines (see Table S1). Right panel: one-dimensional cuts through the contribution function for the highest-frequency ($J_K=4_3-3_3$) PH$_3$ line, at various frequency offsets from the line core frequency ($f_0=1067.2063$~GHz). \label{fig:cfs}}
\end{figure}

The full spectral contribution function from our 4G2 Venus radiative transfer model is shown in Figure \ref{fig:cfs}. The contribution function includes all species in the model, and reveals that the narrow PH$_3$ $J=4$ spectral line cores only begin to become readily distinguishable from the heavily-broadened line wings for altitudes above 70 km. The right panel of Figure \ref{fig:cfs} shows more clearly the vertical behaviour of the $J_K=4_3-3_3$ line core absorption compared with the wings, with similar amounts of absorption from the line core ($f_0$) and various line wing positions for altitudes $\lesssim70$~km. The $f_0\pm18$~MHz window to which our SOFIA observations are most sensitive corresponds to spectroscopically distinct absorption in the altitude range $\approx75$--110~km. This is significantly higher than the $\sim60$--70~km range probed by the infrared observations of \citeA{enc20}, who obtained $a<5$~ppb; our SOFIA observations are therefore consistent with that study, but provide a complementary, stricter upper limit at higher altitudes.

Our 0.8~ppb upper limit is also consistent with \citeA{tro21}, who obtained phosphine upper limits less than 20 ppb from various Venus Express solar occultation measurements in the altitude range 60--95~km. Their strictest upper limits ($\lesssim1$~ppb) were obtained at latitudes within about 20 degrees of Venus's poles, and in the altitude range 60--77~km; their lowest value being 0.2 ppb at 69 km. Consequently, our SOFIA observations, by virtue of being more sensitive to equatorial latitudes, as well as higher altitudes, are again, highly complementary to these previous observations.

We also generated contribution functions for the PH$_3$ $J=1$--0 line (the same one observed by \citeA{gre21a} using ALMA and JCMT), as shown in figure S4. Due the lower energy of the $J=1$ level ($E_u=13$~K, \emph{vs.} 121~K for $J=4_3$), the \citeA{gre21a} observations sound slightly higher altitudes, a few kilometers above our SOFIA observations. Figure S4 (right panel) also shows that the continuum-subtracted spectral region within $\pm5$~km\,s$^{-1}$ ($\pm4.5$~MHz) of the PH$_3$ $J=1$--0 line is only sensitive to emission from above 75~km. This is significantly higher than the altitude of Venus's cloud tops (around 60--70~km), and is at odds with the claim by \citeA{gre21a} that they detected PH$_3$ in the cloud decks.

The non-detection of PH$_3$ by several teams confirms a strict upper limit on the phosphine gas abundance across a range of latitudes within the mid-to-upper atmosphere of Venus. Conversely, strong evidence for the presence of other phosphorous-bearing compounds in Venus's atmosphere was provided by X-ray radiometry of aerosol particles by the Vega-1 and Vega-2 descent probes \cite{and87a}. Surprisingly, in the altitude range 47--52~km, the cloud aerosols were found to consist primarily of P-bearing compounds, while between 52--62~km, they were dominated by sulfur- and chlorine-containing species. Based on lithospheric chemical abundance arguments, \citeA{and87b} therefore deduced that phosphorous oxides and phosphoric acids were a likely component of the Venusian cloud particles, plausibly generated from the condensation of volcanic vapors. The participation of phosphorous oxides (P$_n$O$_m$) in atmospheric chemical processes in the cloud layers, including reaction with H$_2$SO$_4$ and other sulphur-bearing species in the presence of H$_2$O, can result in the production of phosphoric acid and HPO-bearing polymers \cite{and87b,kra89}. A complex chemical network involving phosphorous on Venus is therefore in effect. However, the plausibility of forming PH$_3$ as a result of Venus photochemistry was explored by \citeA{bai21}, and no efficient chemical pathways have yet been found. The possible production of detectable quantities of PH$_3$ in and above the clouds, as a result of phosphide release from volcanic plumes \cite{tru21}, also remains questionable \cite{bai22}, so the non-detection of PH$_3$ gas is not surprising given our current knowledge.

\section{Conclusions}

SOFIA GREAT spectra of the PH$_3$ $J=2$--1 and 4--3 transitions obtained toward Venus on 2021 November 10th--13th show no significant evidence for the presence of phosphine gas. From our most sensitive observations (in the 4G2 channel), we derived an upper limit on the PH$_3$ abundance of 0.8~ppb in the altitude range $\approx75$--110~km. The $25''$ telescope beam, pointing at the Venus opto-center, was sensitive to emission from across the entire disk of Venus, but with a peak sensitivity in the vicinity of the dayside equator. As a result of our high spectral-resolution observations, combined with previous infrared studies using Venus Express and IRTF, we now have in-hand a robust body of evidence demonstrating a lack of phosphine gas in Venus's atmosphere, spanning a broad range of altitudes $\sim60-110$~km, with latitudinal coverage ranging from equatorial to polar.

Our measurements, obtained around a single epoch, do not rule out the possibility of time-variability in the Venus PH$_3$ abundance. Due to its utility as a possible indicator for unexpected geological, atmospheric, and biological processes, there is merit in continued searches for PH$_3$ in the atmospheres of Venus and other planets. Further theoretical and laboratory-based investigations of phosphorous chemistry in solid, liquid and vapor phases will also be required, to better understand the planetary sources and sinks of this biologically important element.

%%%%%%%%%%%%%%%%%%%%%%%%%%%%%%%%
%% Optional Appendix goes here
%
% The \appendix command resets counters and redefines section heads
%
% After typing \appendix
%
%\section{Here Is Appendix Title}
% will show
% A: Here Is Appendix Title
%
%\appendix
%\section{Here is a sample appendix}

\section{Open Research}
All observational data used in this study are freely available from the SOFIA science archive at https://irsa.ipac.caltech.edu/applications/sofia using AOR ID 75\_0059\_1. 

Data reduction and analysis scripts (written in Python 3) are accessible from the Zenodo repository \cite{zen22}.

The Planetary Spectrum Generator (PSG) used for producing our spectral models is accessible at psg.gsfc.nasa.gov.

%It is important to cite individual datasets in this section and, and they must be included in your bibliography. Please use the type field in your bibtex file to specify the type of data cited. Options include [Dataset], [Software], [ComputationalNotebook], [Collection].
%Example:
%
%@misc{https://doi.org/10.7283/633e-1497,
%  doi = {10.7283/633E-1497},
%  url = {https://www.unavco.org/data/doi/10.7283/633E-1497},
%  author = {de Zeeuw-van Dalfsen, Elske and Sleeman, Reinoud},
%  title = {KNMI Dutch Antilles GPS Network - SAB1-St_Johns_Saba_NA P.S.},
%  publisher = {UNAVCO, Inc.},
%  year = {2019},
%  type = {dataset}
%}

%%%%%%%%%%%%%%%%%%%%%%%%%%%%%%%%%%%%%%%%%%%%%%%

\acknowledgments
This work is based on observations made with the NASA/DLR Stratospheric Observatory for Infrared Astronomy (SOFIA). SOFIA is jointly operated by the Universities Space Research Association, Inc. (USRA), under NASA contract NNA17BF53C, and the Deutsches SOFIA Institut (DSI) under DLR contract 50 OK 2002 to the University of Stuttgart. Financial support for this work was provided by NASA through award \#75\_0059 issued by USRA. Research at NASA GSFC was also supported by the NASA Planetary Science Division Internal Scientist Funding Program through the Fundamental Laboratory Research work package (FLaRe). The German Receiver for Astronomy at Terahertz Frequencies GREAT \cite{ris18} is a development by the MPI f{\"u}r Radioastronomie and the KOSMA Universit{\"a}t zu K{\"o}ln, in cooperation with the MPI f{\"u}rSonnensystemforschung and the DLR Institut für Planetenforschung. We gratefully acknowledge support from the GREAT team in setting up and performing the observations.

%% ------------------------------------------------------------------------ %%
%% References and Citations

%%%%%%%%%%%%%%%%%%%%%%%%%%%%%%%%%%%%%%%%%%%%%%%
%
% \bibliography{<name of your .bib file>} don't specify the file extension
%
% don't specify bibliographystyle

% In the References section, cite the data/software described in the Availability Statement (this includes primary and processed data used for your research). For details on data/software citation as well as examples, see the Data & Software Citation section of the Data & Software for Authors guidance
% https://www.agu.org/Publish-with-AGU/Publish/Author-Resources/Data-and-Software-for-Authors#citation

%%%%%%%%%%%%%%%%%%%%%%%%%%%%%%%%%%%%%%%%%%%%%%%

\nocite{pic98} 
\bibliography{venus}

\begin{thebibliography}{}

\bibitem [\protect \citeauthoryear {%
{Akins}%
, {Lincowski}%
, {Meadows}%
\BCBL {}\ \BBA {} {Steffes}%
}{%
{Akins}%
\ \protect \BOthers {.}}{%
{\protect \APACyear {2021}}%
}]{%
aki21}
\APACinsertmetastar {%
aki21}%
\begin{APACrefauthors}%
{Akins}, A\BPBI B.%
, {Lincowski}, A\BPBI P.%
, {Meadows}, V\BPBI S.%
\BCBL {}\ \BBA {} {Steffes}, P\BPBI G.%
\end{APACrefauthors}%
\unskip\
\newblock
\APACrefYearMonthDay{2021}{{\APACmonth{02}}}{}.
\newblock
{\BBOQ}\APACrefatitle {{Complications in the ALMA Detection of Phosphine at
  Venus}} {{Complications in the ALMA Detection of Phosphine at Venus}}.{\BBCQ}
\newblock
\APACjournalVolNumPages{ApJL}{907}{2}{L27}.
\newblock
\begin{APACrefDOI} \doi{10.3847/2041-8213/abd56a} \end{APACrefDOI}
\PrintBackRefs{\CurrentBib}

\bibitem [\protect \citeauthoryear {%
{Andreichikov}%
}{%
{Andreichikov}%
}{%
{\protect \APACyear {1987}}%
}]{%
and87b}
\APACinsertmetastar {%
and87b}%
\begin{APACrefauthors}%
{Andreichikov}, B\BPBI M.%
\end{APACrefauthors}%
\unskip\
\newblock
\APACrefYearMonthDay{1987}{{\APACmonth{09}}}{}.
\newblock
{\BBOQ}\APACrefatitle {{Chemical composition and structure of Venus clouds from
  results of X-ray radiometric experiments made with the Vega 1 and Vega 2
  automatic interplanetary stations.}} {{Chemical composition and structure of
  Venus clouds from results of X-ray radiometric experiments made with the Vega
  1 and Vega 2 automatic interplanetary stations.}}{\BBCQ}
\newblock
\APACjournalVolNumPages{Kosmicheskie Issledovaniia}{25}{}{737-743}.
\PrintBackRefs{\CurrentBib}

\bibitem [\protect \citeauthoryear {%
{Andreichikov}%
\ \protect \BOthers {.}}{%
{Andreichikov}%
\ \protect \BOthers {.}}{%
{\protect \APACyear {1987}}%
}]{%
and87a}
\APACinsertmetastar {%
and87a}%
\begin{APACrefauthors}%
{Andreichikov}, B\BPBI M.%
, Akhmetshin, I\BPBI K.%
, Korchuganov, B\BPBI N.%
, Mukhin, L\BPBI M.%
, Ogorodnikov, B\BPBI I.%
, Petryanov, I\BPBI V.%
\BCBL {}\ \BBA {} Skitovich, V\BPBI I.%
\end{APACrefauthors}%
\unskip\
\newblock
\APACrefYearMonthDay{1987}{{\APACmonth{09}}}{}.
\newblock
{\BBOQ}\APACrefatitle {{Chemical composition and structure of Venus clouds from
  results of X-ray radiometric experiments made with the Vega 1 and Vega 2
  automatic interplanetary stations.}} {{Chemical composition and structure of
  Venus clouds from results of X-ray radiometric experiments made with the Vega
  1 and Vega 2 automatic interplanetary stations.}}{\BBCQ}
\newblock
\APACjournalVolNumPages{Kosmicheskie Issledovaniia}{25}{}{721-736}.
\PrintBackRefs{\CurrentBib}

\bibitem [\protect \citeauthoryear {%
{Bains}%
\ \protect \BOthers {.}}{%
{Bains}%
\ \protect \BOthers {.}}{%
{\protect \APACyear {2021}}%
{\protect \APACexlab {{\protect \BCnt {1}}}}}]{%
bai20}
\APACinsertmetastar {%
bai20}%
\begin{APACrefauthors}%
{Bains}, W.%
, {Petkowski}, J\BPBI J.%
, {Seager}, S.%
, {Ranjan}, S.%
, {Sousa-Silva}, C.%
, {Rimmer}, P\BPBI B.%
\BDBL {}{Richards}, A\BPBI M\BPBI S.%
\end{APACrefauthors}%
\unskip\
\newblock
\APACrefYearMonthDay{2021{\protect \BCnt {1}}}{{\APACmonth{10}}}{}.
\newblock
{\BBOQ}\APACrefatitle {{Phosphine on Venus Cannot Be Explained by Conventional
  Processes}} {{Phosphine on Venus Cannot Be Explained by Conventional
  Processes}}.{\BBCQ}
\newblock
\APACjournalVolNumPages{Astrobiology}{21}{10}{1277-1304}.
\newblock
\begin{APACrefDOI} \doi{10.1089/ast.2020.2352} \end{APACrefDOI}
\PrintBackRefs{\CurrentBib}

\bibitem [\protect \citeauthoryear {%
{Bains}%
\ \protect \BOthers {.}}{%
{Bains}%
\ \protect \BOthers {.}}{%
{\protect \APACyear {2021}}%
{\protect \APACexlab {{\protect \BCnt {2}}}}}]{%
bai21}
\APACinsertmetastar {%
bai21}%
\begin{APACrefauthors}%
{Bains}, W.%
, {Petkowski}, J\BPBI J.%
, {Seager}, S.%
, {Ranjan}, S.%
, {Sousa-Silva}, C.%
, {Rimmer}, P\BPBI B.%
\BDBL {}{Richards}, A\BPBI M\BPBI S.%
\end{APACrefauthors}%
\unskip\
\newblock
\APACrefYearMonthDay{2021{\protect \BCnt {2}}}{{\APACmonth{10}}}{}.
\newblock
{\BBOQ}\APACrefatitle {{Phosphine on Venus Cannot Be Explained by Conventional
  Processes}} {{Phosphine on Venus Cannot Be Explained by Conventional
  Processes}}.{\BBCQ}
\newblock
\APACjournalVolNumPages{Astrobiology}{21}{10}{1277-1304}.
\newblock
\begin{APACrefDOI} \doi{10.1089/ast.2020.2352} \end{APACrefDOI}
\PrintBackRefs{\CurrentBib}

\bibitem [\protect \citeauthoryear {%
{Bains}%
\ \protect \BOthers {.}}{%
{Bains}%
\ \protect \BOthers {.}}{%
{\protect \APACyear {2022}}%
}]{%
bai22}
\APACinsertmetastar {%
bai22}%
\begin{APACrefauthors}%
{Bains}, W.%
, {Shorttle}, O.%
, {Ranjan}, S.%
, {Rimmer}, P\BPBI B.%
, {Petkowski}, J\BPBI J.%
, {Greaves}, J\BPBI S.%
\BCBL {}\ \BBA {} {Seager}, S.%
\end{APACrefauthors}%
\unskip\
\newblock
\APACrefYearMonthDay{2022}{{\APACmonth{02}}}{}.
\newblock
{\BBOQ}\APACrefatitle {{Only extraordinary volcanism can explain the presence
  of parts per billion phosphine on Venus}} {{Only extraordinary volcanism can
  explain the presence of parts per billion phosphine on Venus}}.{\BBCQ}
\newblock
\APACjournalVolNumPages{Proceedings of the National Academy of
  Science}{119}{7}{e2121702119}.
\newblock
\begin{APACrefDOI} \doi{10.1073/pnas.2121702119} \end{APACrefDOI}
\PrintBackRefs{\CurrentBib}

\bibitem [\protect \citeauthoryear {%
Cordiner%
}{%
Cordiner%
}{%
{\protect \APACyear {2022}}%
}]{%
zen22}
\APACinsertmetastar {%
zen22}%
\begin{APACrefauthors}%
Cordiner, M.%
\end{APACrefauthors}%
\unskip\
\newblock
\APACrefYearMonthDay{2022}{{\APACmonth{08}}}{}.
\newblock
{\BBOQ}\APACrefatitle {{Data analysis scripts for submitted article ``Phosphine
  in the Venusian Atmosphere: A Strict Upper Limit from SOFIA GREAT
  Observations'' [Software]}} {{Data analysis scripts for submitted article
  ``Phosphine in the Venusian Atmosphere: A Strict Upper Limit from SOFIA GREAT
  Observations'' [Software]}}.{\BBCQ}
\newblock
\APACjournalVolNumPages{Zenodo}{}{}{}.
\newblock
\begin{APACrefURL} \url{https://doi.org/10.5281/zenodo.7020418}
  \end{APACrefURL}
\newblock
\begin{APACrefDOI} \doi{10.5281/zenodo.7020418} \end{APACrefDOI}
\PrintBackRefs{\CurrentBib}

\bibitem [\protect \citeauthoryear {%
{Duran}%
\ \protect \BOthers {.}}{%
{Duran}%
\ \protect \BOthers {.}}{%
{\protect \APACyear {2021}}%
}]{%
dur21}
\APACinsertmetastar {%
dur21}%
\begin{APACrefauthors}%
{Duran}, C\BPBI A.%
, {Gusten}, R.%
, {Risacher}, C.%
, {Gorlitz}, A.%
, {Klein}, B.%
, {Reyes}, N.%
\BDBL {}{Lis}, D\BPBI C.%
\end{APACrefauthors}%
\unskip\
\newblock
\APACrefYearMonthDay{2021}{{\APACmonth{03}}}{}.
\newblock
{\BBOQ}\APACrefatitle {{4GREAT{\textemdash}A Four-Color Receiver for
  High-Resolution Airborne Terahertz Spectroscopy}} {{4GREAT{\textemdash}A
  Four-Color Receiver for High-Resolution Airborne Terahertz
  Spectroscopy}}.{\BBCQ}
\newblock
\APACjournalVolNumPages{IEEE Transactions on Terahertz Science and
  Technology}{11}{2}{194-204}.
\newblock
\begin{APACrefDOI} \doi{10.1109/TTHZ.2020.3042714} \end{APACrefDOI}
\PrintBackRefs{\CurrentBib}

\bibitem [\protect \citeauthoryear {%
{Ehrenreich}%
\ \protect \BOthers {.}}{%
{Ehrenreich}%
\ \protect \BOthers {.}}{%
{\protect \APACyear {2012}}%
}]{%
ehr12}
\APACinsertmetastar {%
ehr12}%
\begin{APACrefauthors}%
{Ehrenreich}, D.%
, {Vidal-Madjar}, A.%
, {Widemann}, T.%
, {Gronoff}, G.%
, {Tanga}, P.%
, {Barth{\'e}lemy}, M.%
\BDBL {}{Arnold}, L.%
\end{APACrefauthors}%
\unskip\
\newblock
\APACrefYearMonthDay{2012}{{\APACmonth{01}}}{}.
\newblock
{\BBOQ}\APACrefatitle {{Transmission spectrum of Venus as a transiting
  exoplanet}} {{Transmission spectrum of Venus as a transiting
  exoplanet}}.{\BBCQ}
\newblock
\APACjournalVolNumPages{Astron. Astrophys}{537}{}{L2}.
\newblock
\begin{APACrefDOI} \doi{10.1051/0004-6361/201118400} \end{APACrefDOI}
\PrintBackRefs{\CurrentBib}

\bibitem [\protect \citeauthoryear {%
{Encrenaz}%
\ \protect \BOthers {.}}{%
{Encrenaz}%
\ \protect \BOthers {.}}{%
{\protect \APACyear {2020}}%
}]{%
enc20}
\APACinsertmetastar {%
enc20}%
\begin{APACrefauthors}%
{Encrenaz}, T.%
, {Greathouse}, T\BPBI K.%
, {Marcq}, E.%
, {Widemann}, T.%
, {B{\'e}zard}, B.%
, {Fouchet}, T.%
\BDBL {}{Sousa-Silva}, C.%
\end{APACrefauthors}%
\unskip\
\newblock
\APACrefYearMonthDay{2020}{{\APACmonth{11}}}{}.
\newblock
{\BBOQ}\APACrefatitle {{A stringent upper limit of the PH$_{3}$ abundance at
  the cloud top of Venus}} {{A stringent upper limit of the PH$_{3}$ abundance
  at the cloud top of Venus}}.{\BBCQ}
\newblock
\APACjournalVolNumPages{Astron. Astrophys.}{643}{}{L5}.
\newblock
\begin{APACrefDOI} \doi{10.1051/0004-6361/202039559} \end{APACrefDOI}
\PrintBackRefs{\CurrentBib}

\bibitem [\protect \citeauthoryear {%
{Greaves}%
\ \protect \BOthers {.}}{%
{Greaves}%
\ \protect \BOthers {.}}{%
{\protect \APACyear {2021b}}%
}]{%
gre21b}
\APACinsertmetastar {%
gre21b}%
\begin{APACrefauthors}%
{Greaves}, J\BPBI S.%
, {Richards}, A\BPBI M\BPBI S.%
, {Bains}, W.%
, {Rimmer}, P\BPBI B.%
, {Clements}, D\BPBI L.%
, {Seager}, S.%
\BDBL {}{Fraser}, H\BPBI J.%
\end{APACrefauthors}%
\unskip\
\newblock
\APACrefYearMonthDay{2021b}{{\APACmonth{01}}}{}.
\newblock
{\BBOQ}\APACrefatitle {{Reply to: No evidence of phosphine in the atmosphere of
  Venus from independent analyses}} {{Reply to: No evidence of phosphine in the
  atmosphere of Venus from independent analyses}}.{\BBCQ}
\newblock
\APACjournalVolNumPages{Nature Astronomy}{5}{}{636-639}.
\newblock
\begin{APACrefDOI} \doi{10.1038/s41550-021-01424-x} \end{APACrefDOI}
\PrintBackRefs{\CurrentBib}

\bibitem [\protect \citeauthoryear {%
{Greaves}%
\ \protect \BOthers {.}}{%
{Greaves}%
\ \protect \BOthers {.}}{%
{\protect \APACyear {2021a}}%
}]{%
gre21a}
\APACinsertmetastar {%
gre21a}%
\begin{APACrefauthors}%
{Greaves}, J\BPBI S.%
, {Richards}, A\BPBI M\BPBI S.%
, {Bains}, W.%
, {Rimmer}, P\BPBI B.%
, {Sagawa}, H.%
, {Clements}, D\BPBI L.%
\BDBL {}{Hoge}, J.%
\end{APACrefauthors}%
\unskip\
\newblock
\APACrefYearMonthDay{2021a}{{\APACmonth{01}}}{}.
\newblock
{\BBOQ}\APACrefatitle {{Phosphine gas in the cloud decks of Venus}} {{Phosphine
  gas in the cloud decks of Venus}}.{\BBCQ}
\newblock
\APACjournalVolNumPages{Nature Astronomy}{5}{}{655-664}.
\newblock
\begin{APACrefDOI} \doi{10.1038/s41550-020-1174-4} \end{APACrefDOI}
\PrintBackRefs{\CurrentBib}

\bibitem [\protect \citeauthoryear {%
{Greaves}%
\ \protect \BOthers {.}}{%
{Greaves}%
\ \protect \BOthers {.}}{%
{\protect \APACyear {2022}}%
}]{%
gre22}
\APACinsertmetastar {%
gre22}%
\begin{APACrefauthors}%
{Greaves}, J\BPBI S.%
, {Rimmer}, P\BPBI B.%
, {Richards}, A\BPBI M\BPBI S.%
, {Petkowski}, J\BPBI J.%
, {Bains}, W.%
, {Ranjan}, S.%
\BDBL {}{Fraser}, H\BPBI J.%
\end{APACrefauthors}%
\unskip\
\newblock
\APACrefYearMonthDay{2022}{{\APACmonth{08}}}{}.
\newblock
{\BBOQ}\APACrefatitle {{Low levels of sulphur dioxide contamination of Venusian
  phosphine spectra}} {{Low levels of sulphur dioxide contamination of Venusian
  phosphine spectra}}.{\BBCQ}
\newblock
\APACjournalVolNumPages{MNRAS}{514}{2}{2994-3001}.
\newblock
\begin{APACrefDOI} \doi{10.1093/mnras/stac1438} \end{APACrefDOI}
\PrintBackRefs{\CurrentBib}

\bibitem [\protect \citeauthoryear {%
{Klein}%
\ \protect \BOthers {.}}{%
{Klein}%
\ \protect \BOthers {.}}{%
{\protect \APACyear {2012}}%
}]{%
kle12}
\APACinsertmetastar {%
kle12}%
\begin{APACrefauthors}%
{Klein}, B.%
, {Hochg\"urtel}, S.%
, {Kr\"amer}, I.%
, {Bell}, A.%
, {Meyer}, K.%
\BCBL {}\ \BBA {} {G\"usten}, R.%
\end{APACrefauthors}%
\unskip\
\newblock
\APACrefYearMonthDay{2012}{}{}.
\newblock
{\BBOQ}\APACrefatitle {High-resolution wide-band fast Fourier transform
  spectrometers} {High-resolution wide-band fast fourier transform
  spectrometers}.{\BBCQ}
\newblock
\APACjournalVolNumPages{A\&A}{542}{}{L3}.
\newblock
\begin{APACrefURL} \url{https://doi.org/10.1051/0004-6361/201218864}
  \end{APACrefURL}
\newblock
\begin{APACrefDOI} \doi{10.1051/0004-6361/201218864} \end{APACrefDOI}
\PrintBackRefs{\CurrentBib}

\bibitem [\protect \citeauthoryear {%
{Krasnopolsky}%
}{%
{Krasnopolsky}%
}{%
{\protect \APACyear {1989}}%
}]{%
kra89}
\APACinsertmetastar {%
kra89}%
\begin{APACrefauthors}%
{Krasnopolsky}, V\BPBI A.%
\end{APACrefauthors}%
\unskip\
\newblock
\APACrefYearMonthDay{1989}{{\APACmonth{07}}}{}.
\newblock
{\BBOQ}\APACrefatitle {{Vega mission results and chemical composition of
  Venusian clouds}} {{Vega mission results and chemical composition of Venusian
  clouds}}.{\BBCQ}
\newblock
\APACjournalVolNumPages{Icarus}{80}{1}{202-210}.
\newblock
\begin{APACrefDOI} \doi{10.1016/0019-1035(89)90168-1} \end{APACrefDOI}
\PrintBackRefs{\CurrentBib}

\bibitem [\protect \citeauthoryear {%
{Lincowski}%
\ \protect \BOthers {.}}{%
{Lincowski}%
\ \protect \BOthers {.}}{%
{\protect \APACyear {2021}}%
}]{%
lin21}
\APACinsertmetastar {%
lin21}%
\begin{APACrefauthors}%
{Lincowski}, A\BPBI P.%
, {Meadows}, V\BPBI S.%
, {Crisp}, D.%
, {Akins}, A\BPBI B.%
, {Schwieterman}, E\BPBI W.%
, {Arney}, G\BPBI N.%
\BDBL {}{Domagal-Goldman}, S.%
\end{APACrefauthors}%
\unskip\
\newblock
\APACrefYearMonthDay{2021}{{\APACmonth{02}}}{}.
\newblock
{\BBOQ}\APACrefatitle {{Claimed Detection of PH$_{3}$ in the Clouds of Venus Is
  Consistent with Mesospheric SO$_{2}$}} {{Claimed Detection of PH$_{3}$ in the
  Clouds of Venus Is Consistent with Mesospheric SO$_{2}$}}.{\BBCQ}
\newblock
\APACjournalVolNumPages{ApJL}{908}{2}{L44}.
\newblock
\begin{APACrefDOI} \doi{10.3847/2041-8213/abde47} \end{APACrefDOI}
\PrintBackRefs{\CurrentBib}

\bibitem [\protect \citeauthoryear {%
{Mogul}%
, {Limaye}%
, {Way}%
\BCBL {}\ \BBA {} {Cordova}%
}{%
{Mogul}%
\ \protect \BOthers {.}}{%
{\protect \APACyear {2021}}%
}]{%
mog21}
\APACinsertmetastar {%
mog21}%
\begin{APACrefauthors}%
{Mogul}, R.%
, {Limaye}, S\BPBI S.%
, {Way}, M\BPBI J.%
\BCBL {}\ \BBA {} {Cordova}, J\BPBI A.%
\end{APACrefauthors}%
\unskip\
\newblock
\APACrefYearMonthDay{2021}{{\APACmonth{04}}}{}.
\newblock
{\BBOQ}\APACrefatitle {{Venus' Mass Spectra Show Signs of Disequilibria in the
  Middle Clouds}} {{Venus' Mass Spectra Show Signs of Disequilibria in the
  Middle Clouds}}.{\BBCQ}
\newblock
\APACjournalVolNumPages{GRL}{48}{7}{e91327}.
\newblock
\begin{APACrefDOI} \doi{10.1029/2020GL091327} \end{APACrefDOI}
\PrintBackRefs{\CurrentBib}

\bibitem [\protect \citeauthoryear {%
{Nixon}%
\ \protect \BOthers {.}}{%
{Nixon}%
\ \protect \BOthers {.}}{%
{\protect \APACyear {2010}}%
}]{%
nix10}
\APACinsertmetastar {%
nix10}%
\begin{APACrefauthors}%
{Nixon}, C\BPBI A.%
, {Achterberg}, R\BPBI K.%
, {Teanby}, N\BPBI A.%
, {Irwin}, P\BPBI G\BPBI J.%
, {Flaud}, J\BHBI M.%
, {Kleiner}, I.%
\BDBL {}{Flasar}, F\BPBI M.%
\end{APACrefauthors}%
\unskip\
\newblock
\APACrefYearMonthDay{2010}{{\APACmonth{01}}}{}.
\newblock
{\BBOQ}\APACrefatitle {{Upper limits for undetected trace species in the
  stratosphere of Titan}} {{Upper limits for undetected trace species in the
  stratosphere of Titan}}.{\BBCQ}
\newblock
\APACjournalVolNumPages{Faraday Discussions}{147}{}{65}.
\newblock
\begin{APACrefDOI} \doi{10.1039/c003771k} \end{APACrefDOI}
\PrintBackRefs{\CurrentBib}

\bibitem [\protect \citeauthoryear {%
{Pickett}%
\ \protect \BOthers {.}}{%
{Pickett}%
\ \protect \BOthers {.}}{%
{\protect \APACyear {1998}}%
}]{%
pic98}
\APACinsertmetastar {%
pic98}%
\begin{APACrefauthors}%
{Pickett}, H\BPBI M.%
, {Poynter}, R\BPBI L.%
, {Cohen}, E\BPBI A.%
, {Delitsky}, M\BPBI L.%
, {Pearson}, J\BPBI C.%
\BCBL {}\ \BBA {} {M{\"u}ller}, H\BPBI S\BPBI P.%
\end{APACrefauthors}%
\unskip\
\newblock
\APACrefYearMonthDay{1998}{{\APACmonth{11}}}{}.
\newblock
{\BBOQ}\APACrefatitle {{Submillimeter, millimeter and microwave spectral line
  catalog.}} {{Submillimeter, millimeter and microwave spectral line
  catalog.}}{\BBCQ}
\newblock
\APACjournalVolNumPages{JQSRT}{60}{5}{883-890}.
\newblock
\begin{APACrefDOI} \doi{10.1016/S0022-4073(98)00091-0} \end{APACrefDOI}
\PrintBackRefs{\CurrentBib}

\bibitem [\protect \citeauthoryear {%
{Risacher}%
\ \protect \BOthers {.}}{%
{Risacher}%
\ \protect \BOthers {.}}{%
{\protect \APACyear {2018}}%
}]{%
ris18}
\APACinsertmetastar {%
ris18}%
\begin{APACrefauthors}%
{Risacher}, C.%
, {G{\"u}sten}, R.%
, {Stutzki}, J.%
, {H{\"u}bers}, H\BPBI W.%
, {Aladro}, R.%
, {Bell}, A.%
\BDBL {}{Wohler}, B.%
\end{APACrefauthors}%
\unskip\
\newblock
\APACrefYearMonthDay{2018}{{\APACmonth{01}}}{}.
\newblock
{\BBOQ}\APACrefatitle {{The upGREAT Dual Frequency Heterodyne Arrays for
  SOFIA}} {{The upGREAT Dual Frequency Heterodyne Arrays for SOFIA}}.{\BBCQ}
\newblock
\APACjournalVolNumPages{Journal of Astronomical
  Instrumentation}{7}{4}{1840014}.
\newblock
\begin{APACrefDOI} \doi{10.1142/S2251171718400147} \end{APACrefDOI}
\PrintBackRefs{\CurrentBib}

\bibitem [\protect \citeauthoryear {%
{Scargle}%
}{%
{Scargle}%
}{%
{\protect \APACyear {1982}}%
}]{%
sca82}
\APACinsertmetastar {%
sca82}%
\begin{APACrefauthors}%
{Scargle}, J\BPBI D.%
\end{APACrefauthors}%
\unskip\
\newblock
\APACrefYearMonthDay{1982}{{\APACmonth{12}}}{}.
\newblock
{\BBOQ}\APACrefatitle {{Studies in astronomical time series analysis. II.
  Statistical aspects of spectral analysis of unevenly spaced data.}} {{Studies
  in astronomical time series analysis. II. Statistical aspects of spectral
  analysis of unevenly spaced data.}}{\BBCQ}
\newblock
\APACjournalVolNumPages{ApJ}{263}{}{835-853}.
\newblock
\begin{APACrefDOI} \doi{10.1086/160554} \end{APACrefDOI}
\PrintBackRefs{\CurrentBib}

\bibitem [\protect \citeauthoryear {%
{Snellen}%
, {Guzman-Ramirez}%
, {Hogerheijde}%
, {Hygate}%
\BCBL {}\ \BBA {} {van der Tak}%
}{%
{Snellen}%
\ \protect \BOthers {.}}{%
{\protect \APACyear {2020}}%
}]{%
sne20}
\APACinsertmetastar {%
sne20}%
\begin{APACrefauthors}%
{Snellen}, I\BPBI A\BPBI G.%
, {Guzman-Ramirez}, L.%
, {Hogerheijde}, M\BPBI R.%
, {Hygate}, A\BPBI P\BPBI S.%
\BCBL {}\ \BBA {} {van der Tak}, F\BPBI F\BPBI S.%
\end{APACrefauthors}%
\unskip\
\newblock
\APACrefYearMonthDay{2020}{{\APACmonth{12}}}{}.
\newblock
{\BBOQ}\APACrefatitle {{Re-analysis of the 267 GHz ALMA observations of Venus.
  No statistically significant detection of phosphine}} {{Re-analysis of the
  267 GHz ALMA observations of Venus. No statistically significant detection of
  phosphine}}.{\BBCQ}
\newblock
\APACjournalVolNumPages{Astron. Astrophys.}{644}{}{L2}.
\newblock
\begin{APACrefDOI} \doi{10.1051/0004-6361/202039717} \end{APACrefDOI}
\PrintBackRefs{\CurrentBib}

\bibitem [\protect \citeauthoryear {%
{Sousa-Silva}%
\ \protect \BOthers {.}}{%
{Sousa-Silva}%
\ \protect \BOthers {.}}{%
{\protect \APACyear {2020}}%
}]{%
sou20}
\APACinsertmetastar {%
sou20}%
\begin{APACrefauthors}%
{Sousa-Silva}, C.%
, {Seager}, S.%
, {Ranjan}, S.%
, {Petkowski}, J\BPBI J.%
, {Zhan}, Z.%
, {Hu}, R.%
\BCBL {}\ \BBA {} {Bains}, W.%
\end{APACrefauthors}%
\unskip\
\newblock
\APACrefYearMonthDay{2020}{{\APACmonth{02}}}{}.
\newblock
{\BBOQ}\APACrefatitle {{Phosphine as a Biosignature Gas in Exoplanet
  Atmospheres}} {{Phosphine as a Biosignature Gas in Exoplanet
  Atmospheres}}.{\BBCQ}
\newblock
\APACjournalVolNumPages{Astrobiology}{20}{2}{235-268}.
\newblock
\begin{APACrefDOI} \doi{10.1089/ast.2018.1954} \end{APACrefDOI}
\PrintBackRefs{\CurrentBib}

\bibitem [\protect \citeauthoryear {%
{Thompson}%
}{%
{Thompson}%
}{%
{\protect \APACyear {2021}}%
}]{%
tho21}
\APACinsertmetastar {%
tho21}%
\begin{APACrefauthors}%
{Thompson}, M\BPBI A.%
\end{APACrefauthors}%
\unskip\
\newblock
\APACrefYearMonthDay{2021}{{\APACmonth{01}}}{}.
\newblock
{\BBOQ}\APACrefatitle {{The statistical reliability of 267-GHz JCMT
  observations of Venus: no significant evidence for phosphine absorption}}
  {{The statistical reliability of 267-GHz JCMT observations of Venus: no
  significant evidence for phosphine absorption}}.{\BBCQ}
\newblock
\APACjournalVolNumPages{MNRAS}{501}{1}{L18-L22}.
\newblock
\begin{APACrefDOI} \doi{10.1093/mnrasl/slaa187} \end{APACrefDOI}
\PrintBackRefs{\CurrentBib}

\bibitem [\protect \citeauthoryear {%
{Trompet}%
\ \protect \BOthers {.}}{%
{Trompet}%
\ \protect \BOthers {.}}{%
{\protect \APACyear {2021}}%
}]{%
tro21}
\APACinsertmetastar {%
tro21}%
\begin{APACrefauthors}%
{Trompet}, L.%
, {Robert}, S.%
, {Mahieux}, A.%
, {Schmidt}, F.%
, {Erwin}, J.%
\BCBL {}\ \BBA {} {Vandaele}, A\BPBI C.%
\end{APACrefauthors}%
\unskip\
\newblock
\APACrefYearMonthDay{2021}{{\APACmonth{01}}}{}.
\newblock
{\BBOQ}\APACrefatitle {{Phosphine in Venus' atmosphere: Detection attempts and
  upper limits above the cloud top assessed from the SOIR/VEx spectra}}
  {{Phosphine in Venus' atmosphere: Detection attempts and upper limits above
  the cloud top assessed from the SOIR/VEx spectra}}.{\BBCQ}
\newblock
\APACjournalVolNumPages{Astron. Astrophys.}{645}{}{L4}.
\newblock
\begin{APACrefDOI} \doi{10.1051/0004-6361/202039932} \end{APACrefDOI}
\PrintBackRefs{\CurrentBib}

\bibitem [\protect \citeauthoryear {%
Truong%
\ \BBA {} Lunine%
}{%
Truong%
\ \BBA {} Lunine%
}{%
{\protect \APACyear {2021}}%
}]{%
tru21}
\APACinsertmetastar {%
tru21}%
\begin{APACrefauthors}%
Truong, N.%
\BCBT {}\ \BBA {} Lunine, J\BPBI I.%
\end{APACrefauthors}%
\unskip\
\newblock
\APACrefYearMonthDay{2021}{}{}.
\newblock
{\BBOQ}\APACrefatitle {Volcanically extruded phosphides as an abiotic source of
  Venusian phosphine} {Volcanically extruded phosphides as an abiotic source of
  venusian phosphine}.{\BBCQ}
\newblock
\APACjournalVolNumPages{Proceedings of the National Academy of
  Sciences}{118}{29}{e2021689118}.
\newblock
\begin{APACrefURL} \url{https://www.pnas.org/doi/abs/10.1073/pnas.2021689118}
  \end{APACrefURL}
\newblock
\begin{APACrefDOI} \doi{10.1073/pnas.2021689118} \end{APACrefDOI}
\PrintBackRefs{\CurrentBib}

\bibitem [\protect \citeauthoryear {%
{Villanueva}%
\ \protect \BOthers {.}}{%
{Villanueva}%
\ \protect \BOthers {.}}{%
{\protect \APACyear {2021}}%
}]{%
vil21}
\APACinsertmetastar {%
vil21}%
\begin{APACrefauthors}%
{Villanueva}, G\BPBI L.%
, {Cordiner}, M.%
, {Irwin}, P\BPBI G\BPBI J.%
, {de Pater}, I.%
, {Butler}, B.%
, {Gurwell}, M.%
\BDBL {}{Kopparapu}, R.%
\end{APACrefauthors}%
\unskip\
\newblock
\APACrefYearMonthDay{2021}{{\APACmonth{01}}}{}.
\newblock
{\BBOQ}\APACrefatitle {{No evidence of phosphine in the atmosphere of Venus
  from independent analyses}} {{No evidence of phosphine in the atmosphere of
  Venus from independent analyses}}.{\BBCQ}
\newblock
\APACjournalVolNumPages{Nature Astronomy}{5}{}{631-635}.
\newblock
\begin{APACrefDOI} \doi{10.1038/s41550-021-01422-z} \end{APACrefDOI}
\PrintBackRefs{\CurrentBib}

\bibitem [\protect \citeauthoryear {%
{Villanueva}%
, {Smith}%
, {Protopapa}%
, {Faggi}%
\BCBL {}\ \BBA {} {Mandell}%
}{%
{Villanueva}%
\ \protect \BOthers {.}}{%
{\protect \APACyear {2018}}%
}]{%
vil18}
\APACinsertmetastar {%
vil18}%
\begin{APACrefauthors}%
{Villanueva}, G\BPBI L.%
, {Smith}, M\BPBI D.%
, {Protopapa}, S.%
, {Faggi}, S.%
\BCBL {}\ \BBA {} {Mandell}, A\BPBI M.%
\end{APACrefauthors}%
\unskip\
\newblock
\APACrefYearMonthDay{2018}{{\APACmonth{09}}}{}.
\newblock
{\BBOQ}\APACrefatitle {{Planetary Spectrum Generator: An accurate online
  radiative transfer suite for atmospheres, comets, small bodies and
  exoplanets}} {{Planetary Spectrum Generator: An accurate online radiative
  transfer suite for atmospheres, comets, small bodies and exoplanets}}.{\BBCQ}
\newblock
\APACjournalVolNumPages{JQSRT}{217}{}{86-104}.
\newblock
\begin{APACrefDOI} \doi{10.1016/j.jqsrt.2018.05.023} \end{APACrefDOI}
\PrintBackRefs{\CurrentBib}

\bibitem [\protect \citeauthoryear {%
{Wunderlich}%
\ \protect \BOthers {.}}{%
{Wunderlich}%
\ \protect \BOthers {.}}{%
{\protect \APACyear {2021}}%
}]{%
wun21}
\APACinsertmetastar {%
wun21}%
\begin{APACrefauthors}%
{Wunderlich}, F.%
, {Scheucher}, M.%
, {Grenfell}, J\BPBI L.%
, {Schreier}, F.%
, {Sousa-Silva}, C.%
, {Godolt}, M.%
\BCBL {}\ \BBA {} {Rauer}, H.%
\end{APACrefauthors}%
\unskip\
\newblock
\APACrefYearMonthDay{2021}{{\APACmonth{03}}}{}.
\newblock
{\BBOQ}\APACrefatitle {{Detectability of biosignatures on LHS 1140 b}}
  {{Detectability of biosignatures on LHS 1140 b}}.{\BBCQ}
\newblock
\APACjournalVolNumPages{Astron. Astrophys.}{647}{}{A48}.
\newblock
\begin{APACrefDOI} \doi{10.1051/0004-6361/202039663} \end{APACrefDOI}
\PrintBackRefs{\CurrentBib}

\bibitem [\protect \citeauthoryear {%
{Zasova}%
, {Moroz}%
, {Linkin}%
, {Khatuntsev}%
\BCBL {}\ \BBA {} {Maiorov}%
}{%
{Zasova}%
\ \protect \BOthers {.}}{%
{\protect \APACyear {2006}}%
}]{%
zas06}
\APACinsertmetastar {%
zas06}%
\begin{APACrefauthors}%
{Zasova}, L\BPBI V.%
, {Moroz}, V\BPBI I.%
, {Linkin}, V\BPBI M.%
, {Khatuntsev}, I\BPBI V.%
\BCBL {}\ \BBA {} {Maiorov}, B\BPBI S.%
\end{APACrefauthors}%
\unskip\
\newblock
\APACrefYearMonthDay{2006}{{\APACmonth{07}}}{}.
\newblock
{\BBOQ}\APACrefatitle {{Structure of the Venusian atmosphere from surface up to
  100 km}} {{Structure of the Venusian atmosphere from surface up to 100
  km}}.{\BBCQ}
\newblock
\APACjournalVolNumPages{Cosmic Research}{44}{4}{364-383}.
\newblock
\begin{APACrefDOI} \doi{10.1134/S0010952506040095} \end{APACrefDOI}
\PrintBackRefs{\CurrentBib}

\end{thebibliography}

%Reference citation instructions and examples:
%
% Please use ONLY \cite and \citeA for reference citations.
% \cite for parenthetical references
% ...as shown in recent studies (Simpson et al., 2019)
% \citeA for in-text citations
% ...Simpson et al. (2019) have shown...
%
%
%...as shown by \citeA{jskilby}.
%...as shown by \citeA{lewin76}, \citeA{carson86}, \citeA{bartoldy02}, and \citeA{rinaldi03}.
%...has been shown \cite{jskilbye}.
%...has been shown \cite{lewin76,carson86,bartoldy02,rinaldi03}.
%... \cite <i.e.>[]{lewin76,carson86,bartoldy02,rinaldi03}.
%...has been shown by \cite <e.g.,>[and others]{lewin76}.
%
% apacite uses < > for prenotes and [ ] for postnotes
% DO NOT use other cite commands (e.g., \citet, \citep, \citeyear, \citealp, etc.).
% \nocite is okay to use to add references from your Supporting Information
%
\addtolength{\hoffset}{-1cm}


\section*{Supplementary material}

# Supporting Information for "Phosphine in the Venusian Atmosphere: A Strict Upper Limit from SOFIA GREAT Observations"

M. A. Cordiner[1,2], G. L. Villanueva[1], H. Wiesemeyer[3], S. N. Milam[1], I. de Pater[4], A. Moullet[5], R. Aladro[3], C. A. Nixon[1], A. E. Thelen[1], S. B. Charnley[1], J. Stutzki[3], V. Kofman[1], S. Faggi[1], G. Liuzzi[1,6], R. Cosentino[7], B. A. McGuire[8]

[1]Solar System Exploration Division, NASA Goddard Space Flight Center, 8800 Greenbelt Road, Greenbelt, MD 20771, USA.

[2]Department of Physics, Catholic University of America, Washington, DC 20064, USA.

[3]Max Planck Institute for Radio Astronomy, Auf dem Hügel 69, D-53121 Bonn, Germany.

[4]Department of Astronomy, University of California, 501 Campbell Hall, Berkeley, CA 94720-3411, USA.

[5]NASA Ames Research Center Moffett Field, CA 94035, USA.

[6]School of Engineering, Università degli Studi della Basilicata, 85100, Potenza, Italy

[7]Space Telescope Science Institute, Baltimore, MD 21218, USA.

[8]Department of Chemistry, Massachusetts Institute of Technology, Cambridge, MA 02139, USA.

## Contents of this file

## Introduction

Table S1 shows the spectral line frequencies, quantum numbers and other spectroscopic parameters for our targeted $PH_3$ lines.

Figures S1 and S2 illustrate the stepwise (iterative) fringe removal process for the SOFIA 4G1 and 4G2 Venus spectra. Successive steps in the spectral fringe removal process, and the corresponding Lomb-Scargle periodogram at each step, are shown.

Figure S3 shows the final impact of our fringe removal process on the Venus $PH_3$ spectral models.

Figure S4 shows contribution functions from our radiative transfer model for the $PH_3$ $J = 1 - 0$ line observed by Greaves et al. (2021a).

Supporting Text S1 describes the origin of the baseline ripples in the calibrated SOFIA 4GREAT spectra.

**Table S1.** Target $PH_3$ lines and spectroscopic parameters

| Freq. (GHz) | $J''_{K''} - J'_{K'}$ | $A$ (s$^{-1}$) | $g_u$ | $E_u$ (K) |
|---|---|---|---|---|
| 533.7946 | $2_0 - 1_0$ | $2.33 \times 10^{-4}$ | 10 | 38.4 |
| 533.8153 | $2_1 - 1_1$ | $1.75 \times 10^{-4}$ | 10 | 37.7 |
| 1066.8359 | $4_0 - 3_0$ | $2.07 \times 10^{-3}$ | 18 | 128.1 |
| 1066.8769 | $4_1 - 3_1$ | $1.94 \times 10^{-3}$ | 18 | 127.3 |
| 1067.0003 | $4_2 - 3_2$ | $1.55 \times 10^{-3}$ | 18 | 125.0 |
| 1067.2063 | $4_3 - 3_3$ | $9.06 \times 10^{-4}$ | 36 | 121.2 |

Ref. — JPL Molecular Spectroscopy catalogue (Pickett et al., 1998).

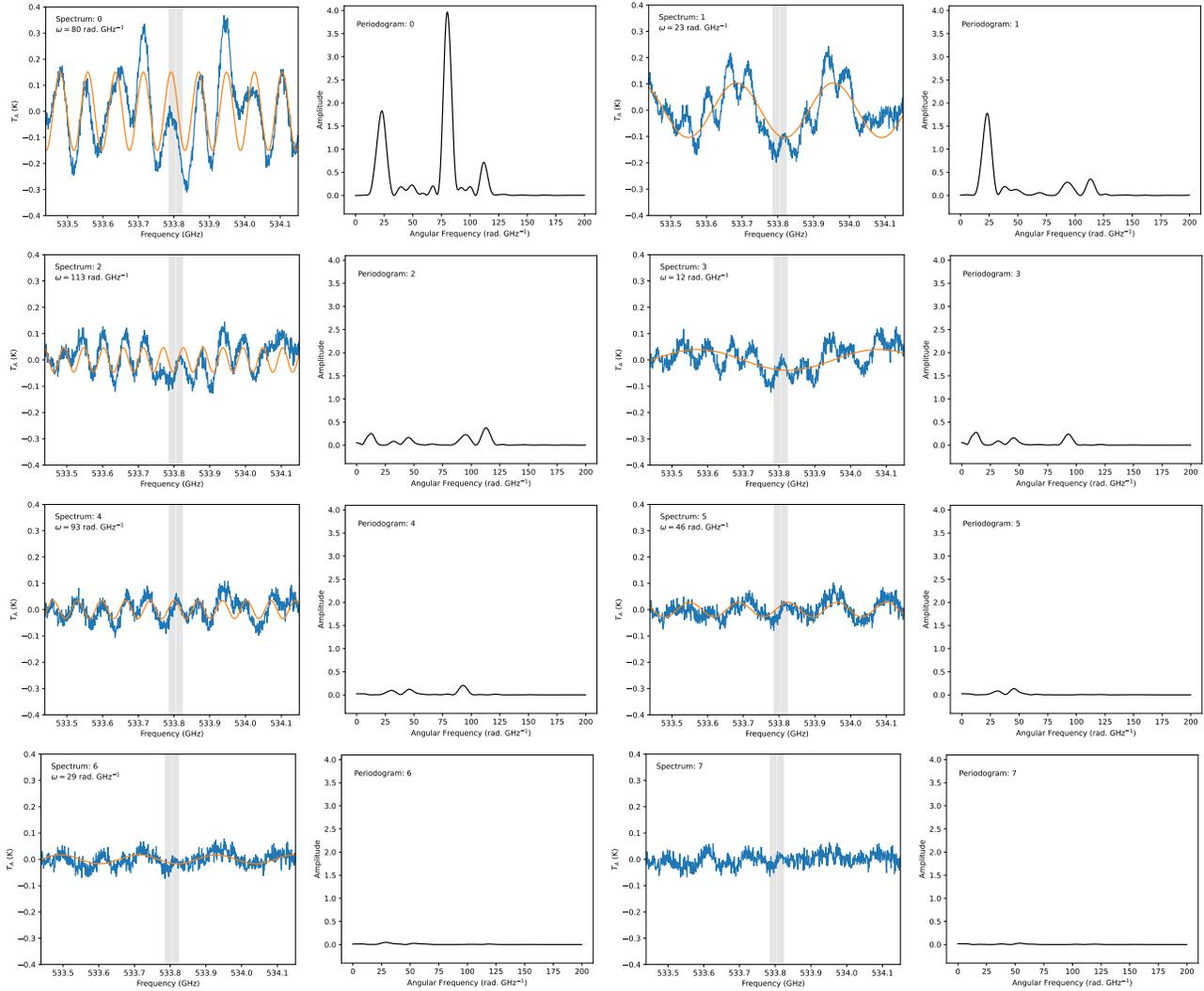

**Figure S1.** Telluric and continuum-subtracted SOFIA 4G1 Venus spectra (columns 1 and 3), and associated Lomb-Scargle periodograms (columns 2 and 4). Each spectrum represents one iteration of our fringe removal method, up to the final spectrum (lower right), which has 7 periodic fringes removed. Grey bars indicate masked regions around the $PH_3$ line positions, which were excluded from the periodogram analysis. For each spectral iteration, the observational data are in blue and the fitted sinusoidal fringe is in orange, the angular frequency of which ($\omega$) is given in the upper left.

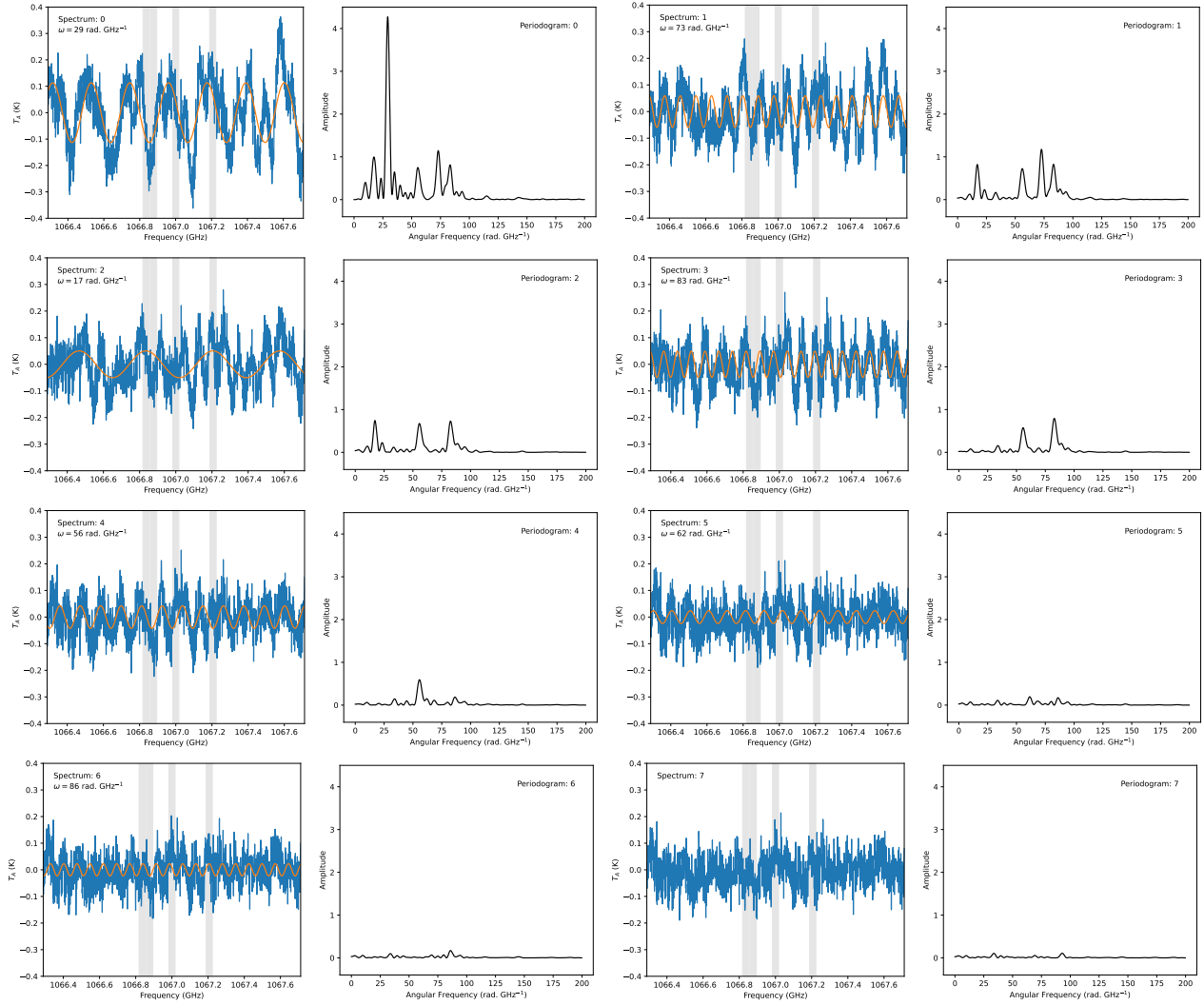

**Figure S2.** Telluric and continuum-subtracted SOFIA 4G2 Venus spectra (columns 1 and 3), and associated Lomb-Scargle periodograms (columns 2 and 4). Each spectrum represents one iteration of our fringe removal method, up to the final spectrum (lower right), which has 7 periodic fringes removed. Grey bars indicate masked regions around the PH$_3$ line positions, which were excluded from the periodogram analysis. For each spectral iteration, the observational data are in blue and the fitted sinusoidal fringe is in orange, the angular frequency of which ($\omega$) is given in the upper left.

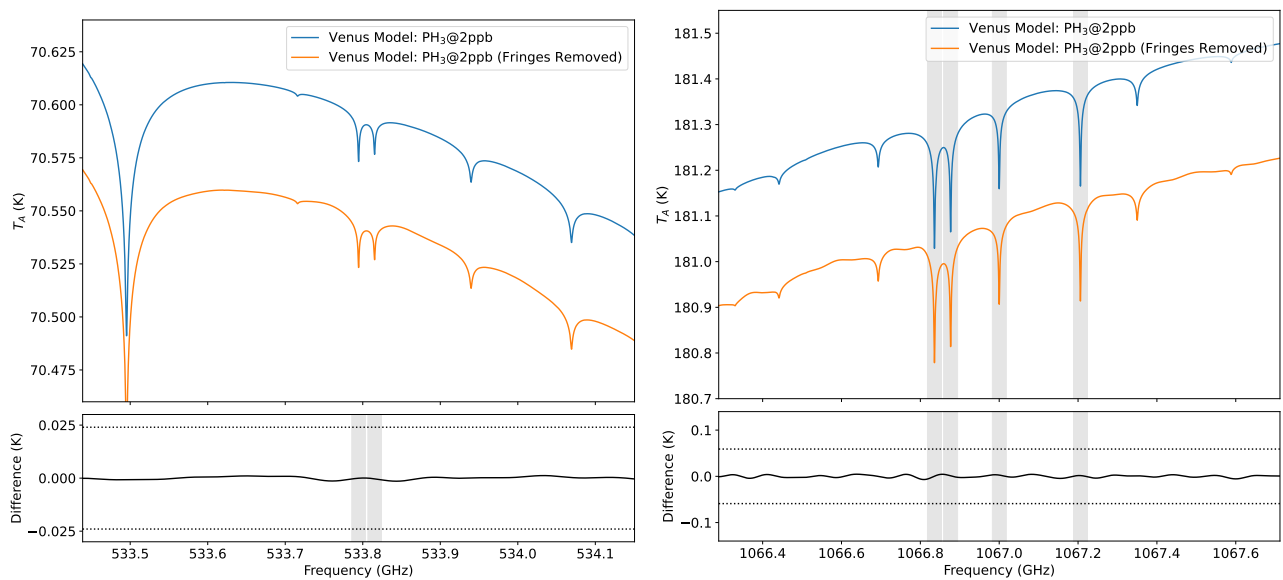

**Figure S3.** Venus PSG spectral models (blue traces) including $PH_3$ at a uniform abundance of 2 ppb for 4G1 (left) and 4G2 (right). The orange traces show the spectra after being subjected to the same defringing method as the observed spectra, and have been shifted vertically for display. Lower panels show the difference spectra ('before' minus 'after' defringing), and in both cases, the residual ripples are well below the RMS noise level, which is indicated with dotted lines.

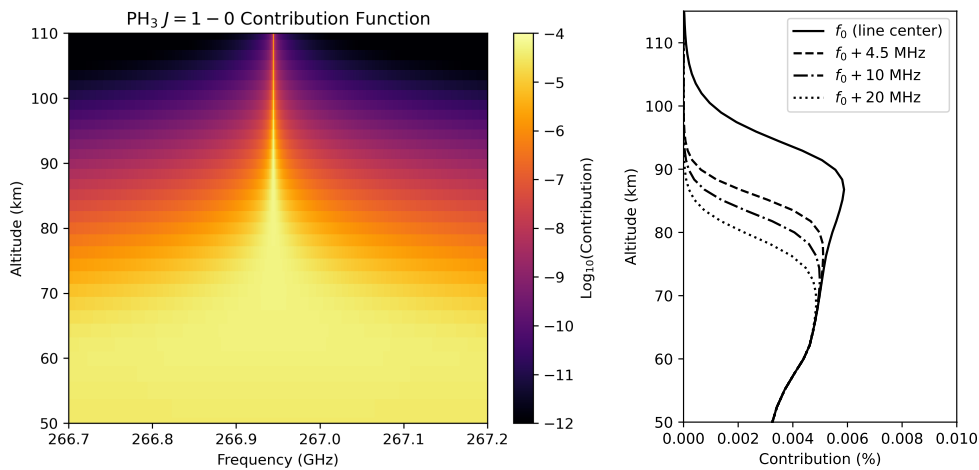

**Figure S4.** Left panel: contribution function from our Venus radiative transfer model, showing the fraction of the total spectral opacity as a function of frequency and altitude for the $PH_3$ $J = 1-0$ line observed by Greaves et al. (2021a). Right panel: one-dimensional cuts through the contribution function, at various frequency offsets from the line core frequency ($f_0$).

## S1. Origin of the baseline continuum ripples

Baseline ripples in the 4G1 and 4G2 channels manifest as a quasi-periodic fringe pattern superimposed on the Venus continuum, with an amplitude of about 0.7% of the continuum intensity (see Figure 1 and Figures S1 and S2). These ripples arise from the presence of standing waves between various optical elements of the SOFIA telescope, *e.g.*, the subreflector, and of the 4GREAT instrument with its mixers. The standing waves are additionally modulated by the frequency-dependent (and sideband-dependent) gain profile of the heterodyne instrument, in both the sky frequency and the intermediate frequency domains. As a result of the double-sideband nature of the receiver, and the fact that the power levels in the signal and image sidebands are not exactly matched, the receiver's response to the standing waves is nonlinear, and consequently, they are not fully removed by the spectral calibration procedure, which is expressed by the following equation:

$$\tilde{T}_A = \frac{\tilde{g}_s \circ (\tilde{P}_s^{on} - \tilde{P}_s^{off}) + \tilde{g}_i \circ (\tilde{P}_i^{on} - \tilde{P}_i^{off})}{\tilde{g}_s \circ (\tilde{P}_s^{hot} - \tilde{P}_s^{cold}) + \tilde{g}_i \circ (\tilde{P}_i^{hot} - \tilde{P}_i^{cold})} (T_R^{hot} - T_R^{cold}) \tag{1}$$

In this equation, spectra are denoted as vectors, with $\tilde{P}$ being the total power count rates in the *on* and *off* source positions, or towards the *hot* and *cold* calibration loads, as indexed. Subscripts $s$ and $i$ refer to the signal and image sidebands, respectively, $\circ$ denotes the element-wise (Hadamard) vector product, and $\tilde{T}_A$ is the resulting calibrated spectrum. For simplicity, the above equation assumes that the Rayleigh-Jeans temperatures of the hot and cold loads are equal in both sidebands. Equal count rates (within the noise) in the signal and image bands would result in a flat bandpass, because the gain factors ($\tilde{g}$) cancel. This is indeed the case for the difference between the two off-source beams A and B (located on either side of the optical axis), because in the average total power from A and B (used as the *off* signal for calibrated Venus spectra) it cancels out. Owing to the strong continuum of Venus and its spectral dependence, the gain factors $\tilde{g}_s, i$ no longer cancel, resulting in different total power baseline shapes in the *on* and *off* positions.

Additional harmonics are introduced into the *on* and *off* standing-wave patterns as a result of folding of the waves from the two receiver sideband spectra, which leads to the presence of multiple periodic components, as manifested in the periodograms shown in Figures S1 and S2. Deconvolution of these overlapping (beating) periodic signals is nontrivial, and therefore, a complete physical interpretation/assignment of the periodogram wave frequencies is beyond the scope of the present study.

## References

Greaves, J. S., Richards, A. M. S., Bains, W., Rimmer, P. B., Sagawa, H., Clements, D. L., ... Hoge, J. (2021a, January). Phosphine gas in the cloud decks of Venus. *Nature Astronomy*, *5*, 655-664. doi: 10.1038/s41550-020-1174-4

Pickett, H. M., Poynter, R. L., Cohen, E. A., Delitsky, M. L., Pearson, J. C., & Müller, H. S. P. (1998, November). Submillimeter, millimeter and microwave spectral line catalog. *JQSRT*, *60*(5), 883-890. doi: 10.1016/S0022-4073(98)00091-0



\end{document}